# Can single crystal X-ray diffraction determine a structure uniquely?

Yihan Shen[1#], Yibin Jiang[2#], Jianhua Lin[1], Cheng Wang[2*], Junliang Sun[1*]

[1]College of Chemistry and Molecular Engineering, BNLMS, Peking University, 292 Chengfu Rd., Beijing, China. [2]State Key Laboratory of Physical Chemistry of Solid Surfaces, College of Chemistry and Chemical Engineering, Xiamen University, 422 Siming South Rd., Xiamen, Fujian, China.

[#]These authors contributed equally to this work. *Corresponding authors.

**The diffraction technique is widely used in the determination of crystal structures and is one of the bases for the modern science and technology[1-4]. All related structure determination methods are based on the assumption that perfect single crystal X-ray diffraction (SXRD) can determine a structure uniquely. But as the structure factor phases are lost in SXRD and even more information is lost in powder X-ray diffraction (PXRD), this assumption is still questionable. In this work, we found that structures with certain characteristic can have its sister structure with exactly the same PXRD or even SXRD pattern. A computer program is developed to search the ICSD database, and about 1000 structures were identified to have this characteristic. The original structure and its sister structures can have different space groups, topologies, crystal systems etc. and some may even have multiple sisters. Our studies indicate that special caution is needed since a structure with reasonable atomic positions and perfect match of**

**experimental diffraction intensities could still be wrong.**

**Background and basic theory**

X-ray diffraction (XRD) is a fundamental technique and is widely applied in the determination of crystal structures in physics, chemistry, geology, pharmaceutical, biology, materials science etc. In this technique, electron density map can be obtained by performing inverse Fourier transform of diffraction intensity together with the structure factor phases.[5] Up to now, there are over one million structures in CSD[6], about 160 thousand structures in PDB[7] and about 210 thousand structures in ICSD[8]. However, in SXRD, only the intensity of the structure factors can be obtained, and phase information is lost. In powder XRD (PXRD), as reflections with the similar crystal plane distance are overlapped, even more information is lost.[9] Practically, a chemically reasonable structure with perfect fitting of experimental diffraction data will be regarded as a correct structure, but theoretically, we cannot rule out the possibility that the correct structure is another structure with the same XRD intensities but different phases.

As diffraction intensities are uniquely determined by atom pairs, a transformation that conserves atom pair can generate XRD-indistinguishable structure. For PXRD, as information of different directions are merged, the directions of atom pairs are not concerned. Based on this principle, we first developed a method which can transform a crystal structure to a new one with identical PXRD pattern. In this method, we choose two point group symmetry operations (**A** and **B**) that do not belong to the original symmetry operations of the structure to divide all atoms in a structure into three parts (P, Q, and R): the part P remains unchanged when operation **A** was applied but will be changed by **B**; the part Q remains unchanged under operation **B** but will be changed by **A**; the part R will be changed by both operations **A** and **B**, but **A** and **B** give the same resulted R'. There may be a substructure in the structure that remains unchanged under either operation **A** or **B**; it can then be merged into any part of P, Q or R. The pair of symmetry operations that meet these requirements for a specific structure can be found using a program to first search all possible symmetry operations that leave at least one atom unchanged. Every pair of these operations was then examined to check if they can divide all atoms in the structure into the three parts. If so, a structure with parts P, Q and R' will have the same PXRD pattern as the original structure with parts P, Q and R, although these two structures may be different from each other. A hypothetic structure is shown in Fig. 1 with each part marked. The reason why these two structures have the same PXRD pattern is that the reflection intensities are determined by atomic pairs in the structure.

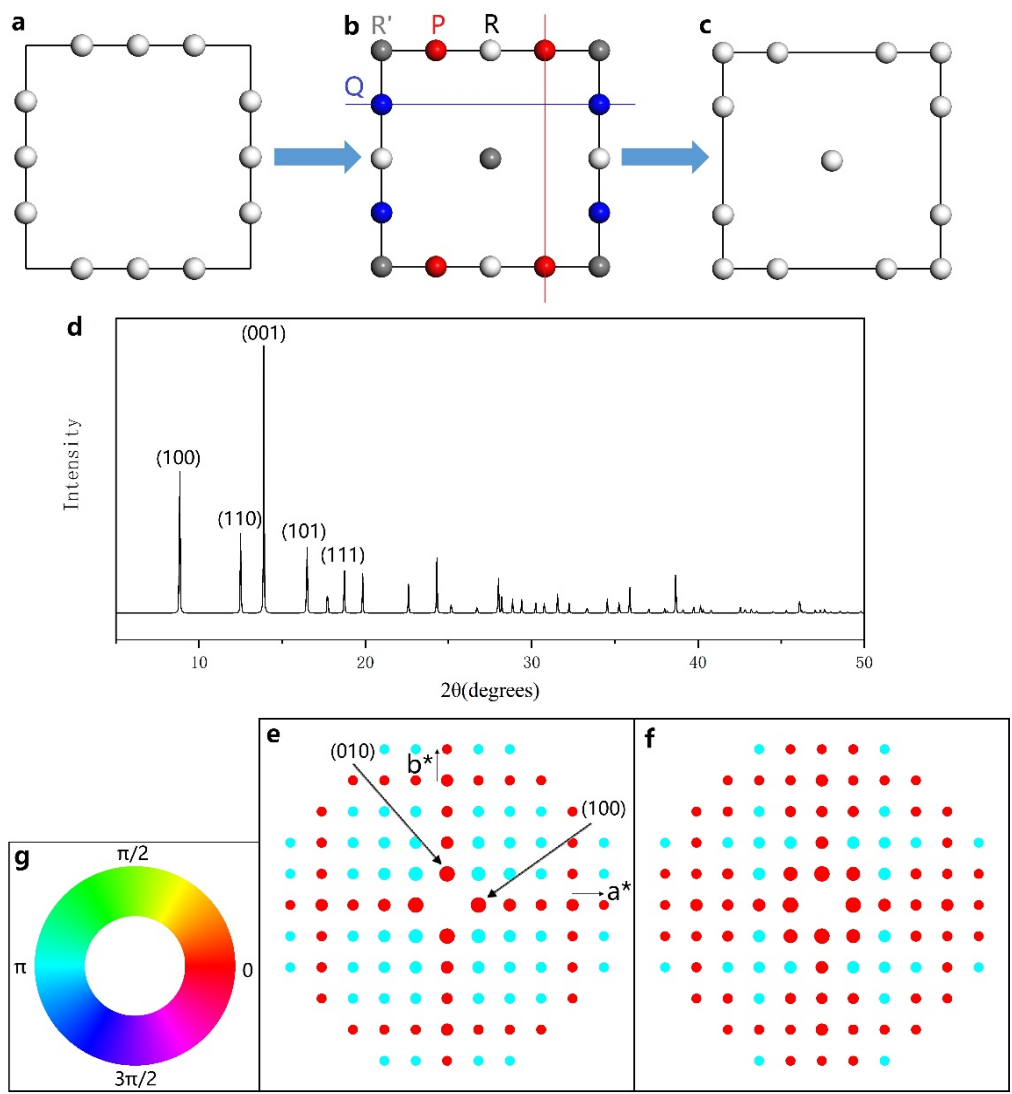

**Figure 1. Hypothetical structure and its sister.** (a) Cell of a hypothetical layered tetragonal structure formed by one type of atom. The *a* and *b*-coordinates for all atoms are 0, 0.25, 0.5 or 0.75 with the *c*-coordinates zero. The lengths of axes are $a=10$ Å and $c=6.37$ Å. (b) The P, Q and R parts of the structure. The red, blue and white spheres represent the part P, Q and R respectively. The red line indicates a mirror symmetry operation **A** exclusive to the part P; the blue line indicates the other mirror **B** exclusive to Q. R' in gray spheres can be generated from R by applying either **A** or **B**. (c) The structure formed by part P, Q and R'. (d) PXRD pattern which is the same for two structures. (e) Structure factor intensities and phases of (hk0) for the structure in (a). (f) Structure factors of (hk0) for the structure in figure (c). (g) Correspondence of phase angle and color in figure (e) and (f).

As each atomic pair in the sister structure (P+Q+R') is generated by **A** or **B** from the original structure (P+Q+R), the distance between them remains the same, and only their orientation is changed. It can be shown that the PXRD intensities of these two structures will be the same.

Note that this is a sufficient condition for two structures to have exactly the same PXRD pattern but not a necessary condition and there may be more PXRD indistinguishable sister structures which cannot be found by this method.

Using this method, a computer program was written and applied to the ICSD database. When considering all atoms as the same element type in each structure, 965 out of 139184 structures have PXRD indistinguishable sister structures, with 866 of them belonging to the space group $Fd\bar{3}m$. Among these structures, 946 structures have SXRD sister structures. When considering the difference of elements, 42 structures are found to have sister structures, among which 31 structures have SXRD sisters.

**XRD sister structures**

As the method discussed above only ensures that two structures are indistinguishable using PXRD, it may find some structure pairs that have only identical PXRD pattern but different SXRD patterns. Using No.29311 FeNiS$_2$[10] in ICSD database as an example, its original structure is in the space group of $Fm\bar{3}m$ (Str. 1a) and its PXRD-indistinguishable sister structure is in $R\bar{3}m$ (Str. 1b). In both structures, Fe and Ni are equally distributed at all four-coordinated metal sites M. In Str. 1a, M$_6$S$_8$ cages are connected by MS$_4$ tetrahedra to form a face-centered cubic structure, while Str. 1b is formed by pure MS layers without connection between layers. In Str. 1a, M atoms in the cage and between the cages have different coordination geometry, but in Str. 1b, all M atoms have identical coordination environment. Here, we use (111) at 15.273º as an example to explain why their PXRD are the same. In Str. 1a, the (-111), (-1-11) and (1-

11) reflections have the same intensity as (111) due to its cubic symmetry. In Str. 1b, (111), (-111), (-1-11) reflections are all absent, while its (1-11) reflection has 4 times intensity, accounting for the PXRD peak at 15.27° of the same intensity as that of Str. 1a.

If the two PXRD indistinguishable sister structures meet some conditions, such as having the same Laue symmetry, they may have identical SXRD patterns and become SXRD indistinguishable. Such two structures may have only minor differences between them, such as No.76432 NaNbO$_3$[11] and its SXRD-indistinguishable sister. Both the original (Str. 2a) and its sister structure (Str. 2b) are distorted Perovskite structures in the space group of $P222_1$ and $P2_12_12_1$ respectively, as shown in Fig. 2a and Fig. 2b. Both structures have two types of octahedral-coordinated Nb atoms with slightly different bond lengths, indicated as blue and cyan spheres in Fig. 2. The difference between the two structures is the ordering of two types of Nb atoms: along z-axis, the same type of Nb atoms exist continuously in Str. 2a but they appear alternatively in Str. 2b. To find out why these two structures have the same SXRD patterns, their reciprocal planes are shown in Fig. 2 together with structure factor phases indicated by colors. For the reflections with $l = 2n$, Str. 2a and Str. 2b are equal in both intensities and phases. For the reflections with $l = 2n+1$, Str. 2a and Str. 2b are equal in intensities, but different in phases. The correlation of the reflections between the two structures can be explained by dividing each structure into two substructures. One substructure which is the same in the two structures contains Na and O in the space group of $I4_122$; the other contains Nb in $C222_1$, which can be related by an inversion center at (0,0,0.125) from Str. 2a to

Str. 2b. As the Na and O substructure and the Nb substructure have different symmetries, the coordination environment of Nb atoms may change after the inversion center operation. For the reflections (h k 2n) which correspond to the reflections from the averaged structure with half c-axis, the intensities and phases are equal since the average structure of Str. 2a and Str. 2b are the same. For the reflections (h k 2n+1), as the Na and O substructure are body-centered, it only contributes to reflections with h+k=2n+1 where h+k+l=2n. The Nb substructure has a C-center which only contributes to reflections with h+k=2n. So when l=2n+1, the Na and O substructure and Nb substructure contribute to different reflections and will not affect each other. Moreover, as Nb substructure only contains one type of element, only phases will change during the inversion center operation even considering anomalous scattering. Consequently, the intensity of reflections with l=2n+1 of Str. 2a and Str. 2b are the same.

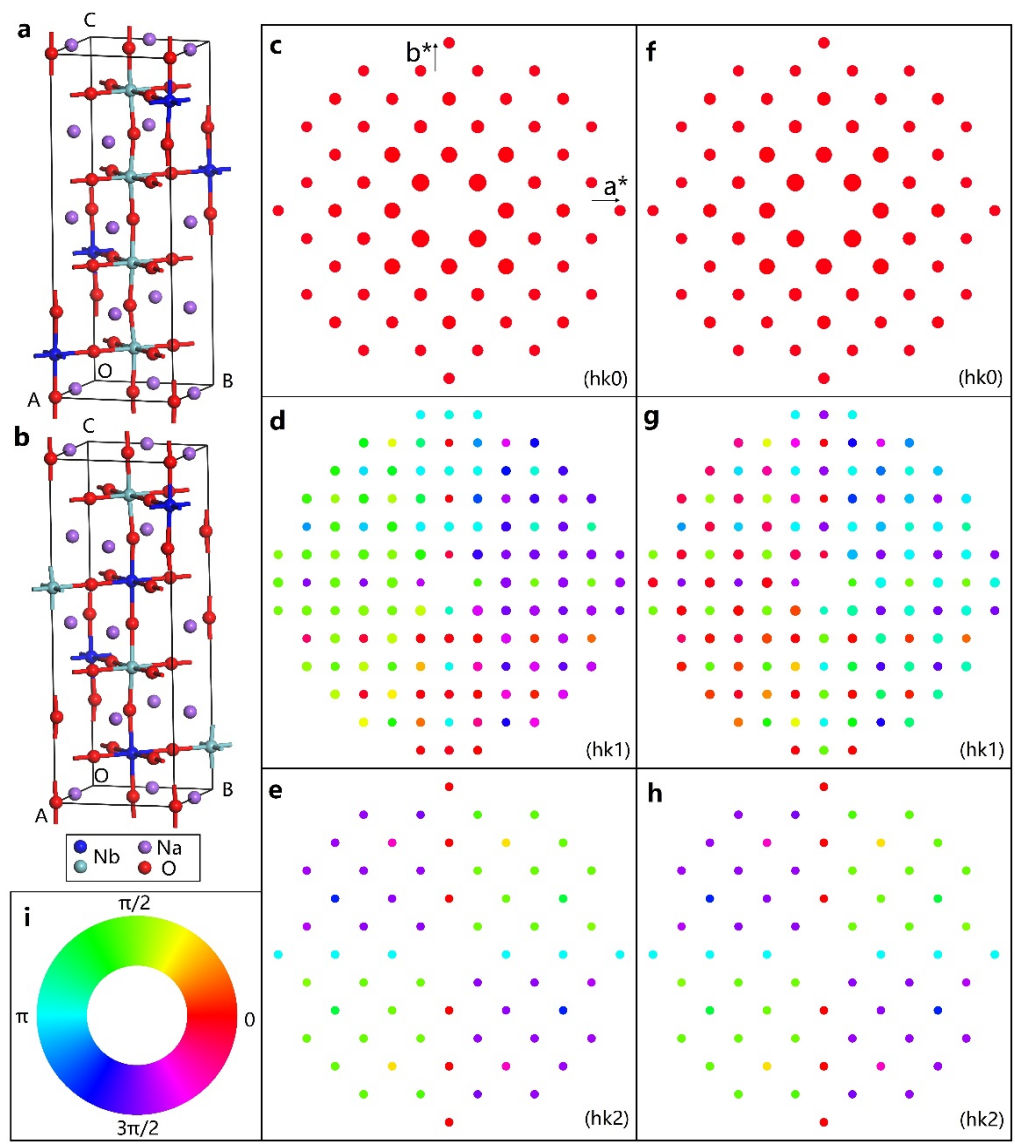

**Figure 2. Two SXRD-indistinguishable structures with small distortions between each other.** (a) Original structure of NaNbO$_3$ (Str. 2a); (b) The sister structure of NaNbO$_3$ (Str. 2b). Nb in blue and cyan present two different distorted octahedral environments, and their ordering is different in two structures. (c)(d)(e) Structure factors of Str. 2a. (f)(g)(h) Structure factors of Str. 2b. (i) Correspondence of phase angle and color in figure (c)~(h). The only difference in reciprocal space is the structure factor phases in (h k 2n+1).

The SXRD-indistinguishable sisters can not only differ in small distortions but also be different in many other aspects, such as coordination geometry, coordination number arrangement, crystal system etc. One example with different coordination geometry is SrPdF$_4$[12] (No.108990) and its indistinguishable sister. The original structure (Str. 3a) and its sister structure (Str. 3b) are shown in Fig. 3. In Str. 3a, Pd has a square-planar coordination geometry, but in Str. 3b, Pd has a tetrahedron coordination geometry. As Pd(II) usually adopts square-planar geometry, Str. 3b is not quite reasonable in a chemical sense but here we only consider the diffraction data to suggest possible physical structure. The structure factors of the two structures differ only by the phases for (h k 2n+1) reflections: the phases are 0 or $\pi$ for the original structure but $\pi/2$ or $-\pi/2$ for the sister structure due to the shift of all F atoms. Interestingly, two SrCrF$_4$ structures with ICSD No.9929[13] and 26105[14] are very similar to Str. 3a and Str. 3b, respectively and the minor difference is that the *z*-coordinate of the only independent F atom is 0.124 here instead of 0.125, making them not perfectly meeting the symmetry restrictions of the three parts discussed above. The tetrahedron coordination structure and square-planar coordination structure thus have slightly different XRD intensities for (h k 2n+1) reflections. For example, the calculated (211), (213) and (215) reflection intensities differ by 3.1%, 9.5% and 16.2% respectively, which could be neglected due to the good fitting of most reflections. As 4-coordinated $Cr^{2+}$ is likely to adopt a square-planar coordination geometry, the square-planar structure is more reasonable for SrCrF$_4$. Another structure (No.24314 TaSe$_2$)[15] in hexagonal symmetry is also found to have similar character with different phases only for (h k 2n+1), where Ta atoms are trigonal prism or octahedral coordinated in two structures as shown in Fig. 3c-d.

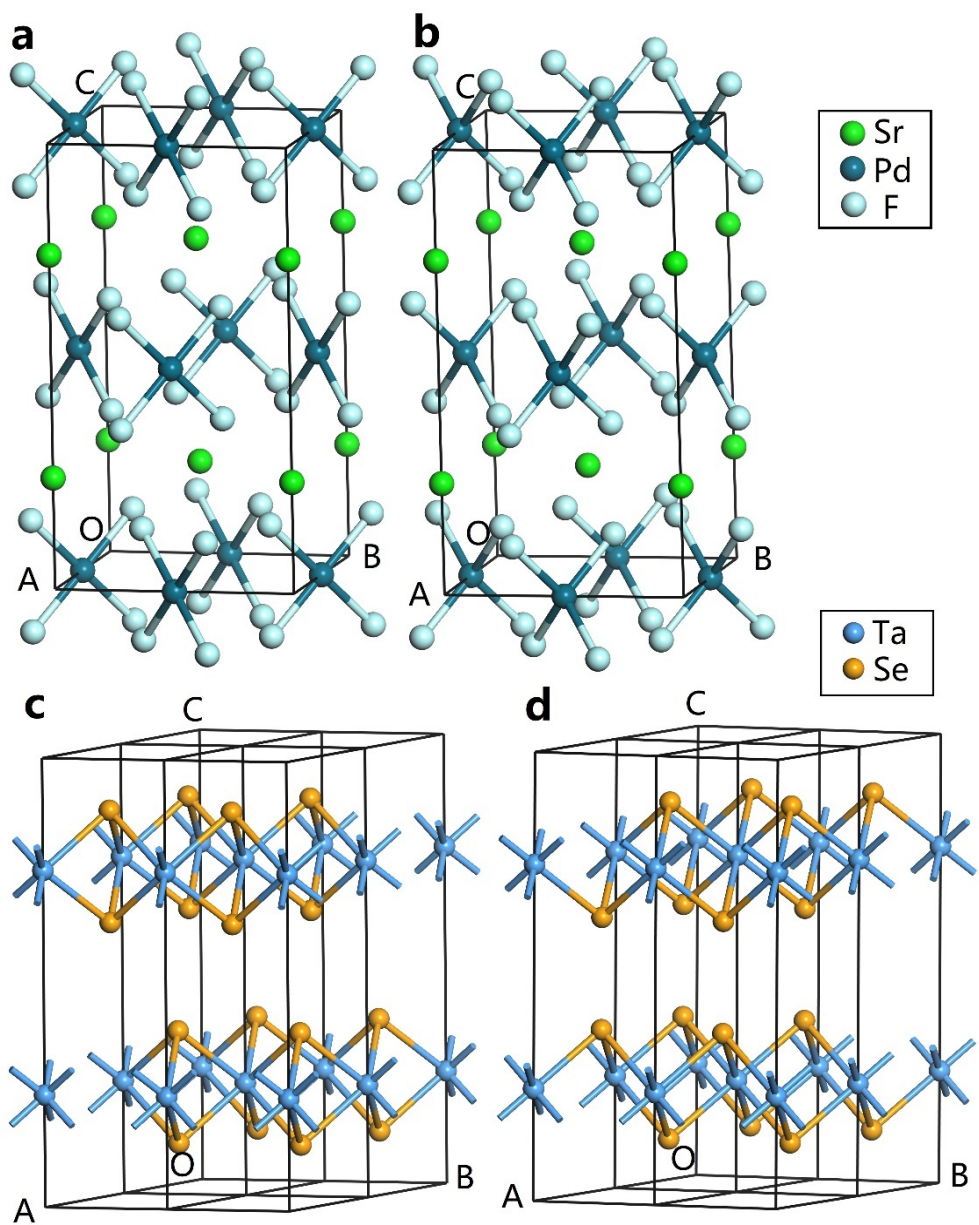

**Figure 3. SXRD-indistinguishable structures with significant coordination geometry changes.** (a) Original structure of SrPdF$_4$ (Str. 3a); (b) The sister structure of SrPdF$_4$ (Str. 3b), where all F atoms are shifted one quarter of the *c*-axis. (c) Original structure of TaSe$_2$; (d) The sister structure of TaSe$_2$, where all Se atoms are shifted one quarter of the *c*-axis.

SXRD sister structures were also found for some complicated structures, such as No.26888 Bi$_{1.5}$Cd$_{0.5}$O$_{2.75}$[16] (Str. 4a) with a SXRD sister (Str. 4b) having different coordination geometry, shown in Fig. 4. Both structures are of $Ia\bar{3}d$ space group with Bi and Cd equally distributed in all metal sites M. The two structures differ in the locations of three O atoms, which experienced a (0,0,0.5) translation from the original structure to its sister structure. The coordination number of M sites in both structures is six, but their disordered octahedral geometries are significantly different, i.e. for the three oxygen atoms related by a 3-fold axis around the M site, the O-M-O angle is 113.58° in Str. 4a and 66.42° in Str. 4b, which resulted in totally different topologies as shown in SI 6. The reciprocal planes of Str. 4a and Str. 4b are shown in Fig. 4, and only some reflections with $l$ odd are different in phases. This can be explained by dividing the O atoms of the Str. 4a into three groups: blue, green and red as indicated in Fig. 4. The three groups are associated with each other by a 3-fold rotation around the body diagonal and adopts tetragonal symmetry by its own with its 4-fold axes along the a, b, c-axis respectively. The blue group experienced a (0,0,0.5) translation from Str. 4a to Str. 4b, which only affects the reflections with l=2n+1. As the blue group has a/2 and (b+c)/2 translation symmetries in the original cell, it only contributes to reflections with h=2n and k+l=2n. So, summing up the scattering power for all three oxygen groups, the three indexes for non-zero reflections should be either all even or two odd and one even; for the latter reflections, the blue group contributes to the reflections with h even, while the green group contributes to the reflections with k even and the red group contributes to the reflections with l even, i.e. when any index is odd, these three oxygen groups contribute to different reflections and will not influence each other. Moreover, the Bi/Cd substructure has the translation symmetry of a/2, b/2 and c/2, and they only contribute to reflections with all indexes even and do not affect others. As a result, the translation of the blue group will only change the phase of (2n 2m+1 2p+1) reflections but not the intensity.

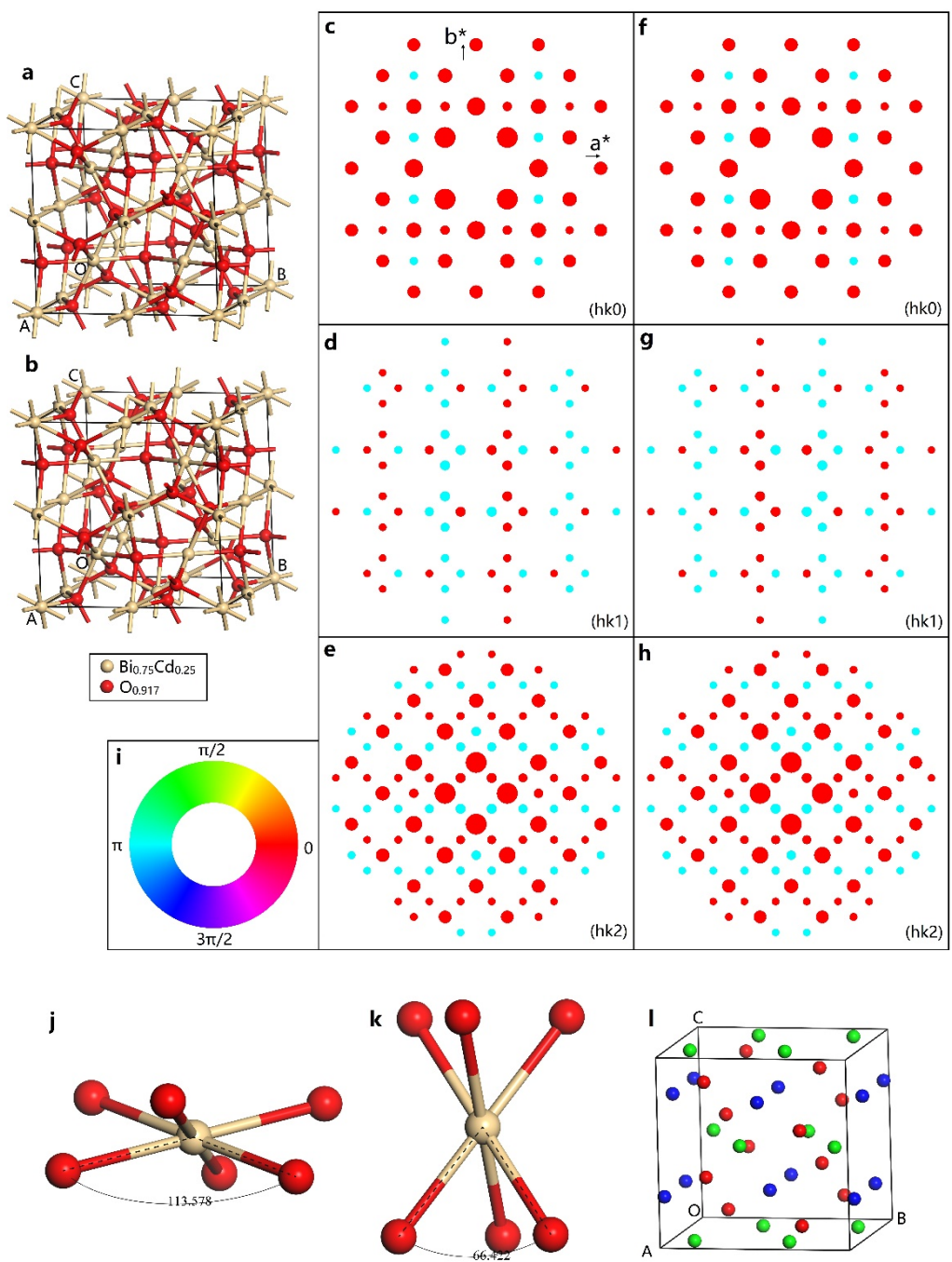

**Figure 4. SXRD-indistinguishable structures with different 3-dimensional topologies and disordered coordination environments.** (a) Original structure of $Bi_{1.5}Cd_{0.5}O_{2.75}$ (Str. 4a); (b) The sister structure of $Bi_{1.5}Cd_{0.5}O_{2.75}$ (Str. 4b). (c)(d)(e) Structure factors of Str. 4a; (f)(g)(h) Strcucture factors of Str. 4b; (i) Correspondence of phase angle and color in figure (c)~(h). (j) The coordination environment of all metal sites in Str. 4a; (k) The coordination environment of all metal sites in Str. 4b. (l) Oxygen atoms of Str. 4a are divided into three groups (blue, red and green) and only the blue

oxygen atoms are shifted in Str. 4b. Since different groups of oxygen contribute diffraction intensities to different reflections, only structure factor phases for the reflections (*h k* 2*n*+1) are different between Str. 4a and Str. 4b.

In some pairs of SXRD sister structures, coordination number rearrangement can be found. No.10364 $K_{0.5}Cr_{0.5}Ni_{0.5}F_6H_2O$[17] is an example. The original structure (Str. 5a) and its sister structure (Str. 5b) are both in $Fd\bar{3}m$ symmetry. Supposing that the atoms at Cr, Ni and K sites have the same scattering power, these two structures will have the same SXRD intensities, and differ in the phases for the (h k 2n+1) reflections. In Str. 5a, K atoms are mainly coordinated by two oxygen atoms with K-O=2.262 Å and Cr/Ni sites are coordinated by six F atoms with Cr/Ni-F=1.941Å. But in Str. 5b, the Cr/Ni/K sites are tetrahedrally coordinated by one O atom and three F atoms with M-O=2.262 Å and M-F=1.941 Å and the coordination environment of O and F is unchanged.

All SXRD sister structure pairs discussed above are of the same crystal system, but SXRD sister structures can also belong to different crystal systems. One example is No.52289 $Cr_7C_3$[18]. This structure (Str. 6a) has a SXRD sister structure (Str. 6b) when anomalous scattering is not considered. Str. 6a is in the symmetry of $P6_3mc$ while Str. 6b is in $Cmc2_1$. The structure change from Str. 6a to Str. 6b can be described as changing the orientations and connections of the $V_3C_4$ groups thus breaking the 6-fold rotation symmetry.

It is also possible that one structure has multiple sister structures. An example is No.18126 $HgI_2$(Str. 7a) [19]. This structure has one SXRD-indistinguishable (Str. 7b) and one PXRD-indistinguishable sister (Str. 7c). The Hg substructure experienced a (0.25,-0.25,-0.25) translation from Str. 7a to Str. 7b, causing reflections with h-k-l=4n+2 and l=2n+1 changing their phases by π, while in Str. 7c, half of the Hg atoms experienced

90º rotation about the 4-fold axis (0.75,0,z) comparing with Str. 7a. In fact, considering the hypothetical structure in Fig. 1a, it has no less than 10 sisters and 4 of them are SXRD-indistinguishable. Thus, sister structures are common, especially for those media-complex structures or those initial structure models with only heavy atoms[20,21].

**Conclusion**

As a standard conclusive technique in modern science, single crystal X-ray diffraction technique is widely used in various area, while in this work, through finding PXRD sister structures of known structures from the database with a general strategy, we found a series of structures which can have their PXRD or even SXRD sister structures, i.e. PXRD/SXRD cannot uniquely determine these structures. Some of these SXRD sister structure pairs only differ slightly in bond length and bond angle, but some pairs differ greatly in coordination geometry, topology, crystal system etc. To distinguish these structures from their sister structures, different complimentary techniques are required for additional information. This result tells that the widely used techniques, PXRD and SXRD, should be used in great caution to avoid acquiring a totally wrong structure, especially when only heavy elements were determined, and in many cases, the wrong initial model with only heavy atoms may prohibit the determination of the entire structure model. This work indicated one possible strategy to find the sister structures which can help in obtaining the correct initial/final model and all obtained crystal structures are recommended to be checked by this method.